\documentclass[5p]{elsarticle}
\usepackage{epsfig}
\usepackage{a4wide}
\usepackage{graphicx,amssymb}
\usepackage{amsmath
}        

\def\lb#1{\if 1#1 \ln\beta \else \ln^#1\beta \fi}
\def\lt#1{\if 1#1 \ln 2 \else \ln^#1 2 \fi}

\newcommand{\be}{\begin{equation}}
\newcommand{\ee}{\end{equation}}
\newcommand{\ba}{\begin{eqnarray}}
\newcommand{\ea}{\end{eqnarray}}

\def\vec#1{{\mbox{\boldmath$#1$}}}

\begin{document}

 \vspace{\baselineskip}

\title{
Top quark spin correlations at the Tevatron and
the LHC
}

    \author{Kirill Melnikov and Markus Schulze}

    \address{
Department of Physics and Astronomy,
Johns Hopkins University,
Baltimore, MD, USA}

    \begin{abstract}
      \noindent
Spin correlations of  top quarks produced
in hadron collisions have not  been observed  experimentally
with large significance. In this Letter, we propose  a new
variable that may enable demonstration of
the existence of spin correlations
with $3-4~\sigma$ significance
using  just a few hundred dilepton events both at the Tevatron and the LHC.
Such number of dilepton events has been  observed at the Tevatron. At the LHC,
it will become available once  integrated  luminosity
of a few hundred inverse picobarns  is collected.
      \end{abstract}

    \maketitle

The existence of spin correlations of
top and anti-top quarks  in $t \bar t$  pair
production in hadron collisions is a solid prediction of
the Standard Model.  The possibility to observe these correlations is unique
to top quarks since  their large masses,
short lifetimes and  the relative weakness
of chromomagnetic fields in the  QCD vacuum,  make it difficult  for
non-perturbative effects to
depolarize $t$ and $ \bar t$ before they
decay. Therefore, if top quarks are
produced in a particular polarization state, spin correlations    can
be observed
by studying kinematic distributions of the top quark decay
products which are sensitive to $t$ and $\bar t$ polarizations.
 For example, in the dilepton channel $pp(p \bar p) \to t \bar t \to
b \bar b \; l \bar l \; \nu \bar \nu$,
 the $V-A$ structure of the
charged current forces momenta of anti-leptons (leptons) to be aligned
(anti-aligned) with
the direction of the top (anti-top) spin vectors.

The traditional way
to study  top quark spin correlations
\cite{Mahlon:1995zn,Stelzer:1995gc,Brandenburg:1996df,Parke:1996pr,Cheung:1996kc,
Mahlon:1997uc,Bernreuther:2000yn,Bernreuther:2001bx,
Bernreuther:2001rq,Bernreuther:2004jv}
is fairly complex.
It involves  choosing   the $t$ and $\bar t$ spin quantization axes
and identifying suitable
reference frames and  angular distributions that are sensitive to
these correlations.  Because of the unobserved
neutrinos in the dilepton events, full kinematics can not
be reconstructed and  determination of quantization  axes and reference
frames
becomes difficult. This feature and a relatively low yield
of dilepton events at the Tevatron  is partially responsible for
the fact that spin correlation
measurements performed by the CDF and D0 collaborations are not conclusive.
For example, the parameter $\kappa$ related to the   top quark
spin asymmetry in the dilepton channel at the Tevatron is predicted
with a very small  uncertainty
in the Standard Model,
$\kappa = 0.78$
\cite{Mahlon:1995zn,Stelzer:1995gc,Brandenburg:1996df,Parke:1996pr,
Mahlon:1997uc,Bernreuther:2000yn,Bernreuther:2001bx,
Bernreuther:2001rq,Bernreuther:2004jv}.  However,
it is measured to be $\kappa = 0.1^{+0.45}_{-0.45}$
and $\kappa = 0.3^{+0.6}_{-0.8}$ by  D0~\cite{d0}
and CDF~\cite{cdf} collaborations, respectively,
 with $5.4(3.0)~{\rm fb}^{-1}$  of integrated luminosity.
Although these results  are
consistent with the Standard Model,
they do not demonstrate the existence of
top quark spin correlations with sufficient significance.
A similar situation occurs when spin correlations are measured
in the lepton plus jets channel~\cite{Aaltonen:2010nz}.

It is then natural to ask
if a better  way exists   to establish the presence  of spin correlations
convincingly.  This question was
recently discussed by G.~Mahlon and S.~Parke
in Ref. \cite{parke}. They
suggested that spin correlations at the LHC
can be observed by measuring the relative  azimuthal
angle $\Delta \phi$ of the two leptons from top decays
{\it in the laboratory frame},
provided that  only events with  the low invariant
mass of $t \bar t$ pairs, $M_{ t \bar t} < 400~{\rm GeV}$,
are accepted.  While in this case it is possible
to distinguish spin-correlated and
spin-uncorrelated events, Ref. \cite{parke} recognizes that placing
a cut on $M_{t \bar t}$ is unphysical since, in dilepton events,
the $t \bar t$ invariant mass can not be fully
reconstructed on an event-by-event basis.
Ref.~\cite{parke} suggests that one can put a cut
on the {\it statistically}
reconstructed invariant mass $M_{t \bar t}$ but
this cut does not seem to work as well  as the
cut on $M_{t \bar t}$ proper.
It was later shown  in Refs.~\cite{schulze,Campbell:2010ff}
that simpler cuts on kinematics of final
state particles -- for example an {\it upper} cut on the transverse momentum
of the charged leptons --
lead to the $\Delta \phi$  laboratory frame
distributions that are sufficiently
different to enable distinguishing  between
spin-correlations and no-spin-corre\-lations hypotheses.

While Ref.~\cite{parke} opened up a new direction in the studies of top
quark spin correlations, similar to previous papers on the subject
\cite{Mahlon:1995zn,Stelzer:1995gc,Brandenburg:1996df,Parke:1996pr,Cheung:1996kc,
Mahlon:1997uc,Bernreuther:2000yn,Bernreuther:2001bx,
Bernreuther:2001rq,Bernreuther:2004jv} it
focused on a single kinematic distribution.
However,  since many
kinematic features of a particular event may be   sensitive
to top quark spin correlations, we should try
to use all the information present in a particular
event to  establish their existence.
To that end,
we ask if, given a set of $t \bar t$ events
observed at the Tevatron or the LHC,
it is possible to distinguish the hypothesis that spins
of the $t \bar t$ pair, entangled in the production process,
remain entangled at the time of their decay, from
the hypothesis that strong QCD dynamics depolarizes produced top
quarks and kinematic features of their decay products are not correlated.

It is well-understood by now
that the  optimal way to answer this question requires the
construction  of a  likelihood function, where
the optimality should be understood in the  sense of
Neyman-Pearson lemma~\cite{np}. In our case, this lemma explains how to
minimize  the probability to accept
the spin-correlation hypothesis when it is false, for
fixed probability to  reject spin-correlation hypothesis when it is
true.   We discuss how to construct the likelihood function.
However, before going into this, we  point out
that we do not consider issues of experimental resolution in this Letter.
In reality, experimental resolution may be  important,
so our results should be considered
as the estimate of  the highest significance with which
two hypotheses  can be separated.

We study dilepton events that contain two neutrinos in the final
state, $ pp (p  \bar p)
\to t \bar t \to b \bar b \; l_1 \bar l_2 \; \bar \nu_{l_1}
\bar \nu_{l_2}$.
Because neutrino momenta can not be measured, the number of measurable
kinematic variables ${\vec x}_{\rm obs}$ is smaller than the
number of kinematic variables
${\vec x} = ({\vec x}_{\rm obs},{\vec x}_{\rm unobs})$
needed to describe a particular event.
The probability distribution for events with
${\vec  x}_{\rm obs}$ set  to a particular value
is computed by integrating  differential cross-sections
for a hypothesis $H$  over
unobserved kinematic variables. We write
\be
P_H({\vec x}_{\rm obs})  =
{\cal N}^{-1}_H\int {\rm d} {\vec x}_{\rm unobs}
\frac{ {\rm d} \sigma_H^{(0)}(\vec x )}{{\rm d} {\vec x}},
\label{eq1}
\ee
where ${\cal N}_H$ is the normalization factor. Given the two hypotheses,
that the top quarks spins are correlated ($H = c$)
or uncorrelated ($H = u$),
we introduce a variable, related to a likelihood ratio for a single event,
that emphasizes the
difference between the two hypotheses
\be
{\cal R}({\vec x}_{\rm obs}) = \frac{P_c({\vec x}_{\rm obs})}{
P_c({\vec x}_{\rm obs})  + P_u({\vec x}_{\rm obs}) }.
\label{eq2a}
\ee
We then  calculate the probability
distribution of the likelihood variable ${\cal R}({\vec x}_{\rm obs})$,
given a particular  hypothesis  about  the underlying physics
\be
\rho_H({\cal R})
= \sigma_H^{-1}\int {\rm d}{\vec x}
\frac{{\rm d} \sigma_H({\vec x})}{{\rm d} {\vec x}}
\delta ( {\cal R}({\vec x}) - {\cal R} ),
\label{eq2}
\ee
and perform statistical tests to see how many events are required to achieve
the  separation of the two hypotheses $H = c,u$.

It is important to realize that,
at the expense of claiming that our likelihood ratio
${\cal R}$ is the {\it optimal} observable~\cite{Cranmer:2006zs}
to separate
spin-correlation and no-spin-correlation hypotheses,
 we can use {\it different cross-sections} to construct
the likelihood variable ${\cal R}({\vec x})$ in Eqs.(\ref{eq1},\ref{eq2a})
and to calculate  the probability distribution
$\rho_H({\cal R})$ in Eq.(\ref{eq2}).
We note  that we use  the {\it Born} differential cross-section
${\rm d} \sigma_H^{(0)}({\vec x})$ to define
${\cal R}({\bf x})$. This is a good choice because
${\rm d} \sigma_H^{(0)}({\vec x})$
captures main kinematic features of the actual physical process and  it
is inexpensive computationally.
However, since this choice does not correspond to the  actual probability
distribution
of the dilepton events, strictly speaking,
${\cal R}$ is not the optimal variable. Nevertheless,
as long as ${\cal R}$
helps  to  separate the two hypotheses, optimality
is not essential.  We emphasize, however, that we
use  the best available approximation to the true
 cross-section ${\rm d} \sigma_H({\vec x})$
to construct the realistic probability
distribution of the variable ${\cal R}$.  To this end,
in this Letter  we employ the next-to-leading order (NLO) QCD prediction
for the top pair production ${\rm d} \sigma_{t \bar t}$ that
includes top quark spin correlations and radiative corrections to
top quark decays~\cite{ms}.

\begin{figure}[t]
\begin{center}
\includegraphics[angle=-90,scale=0.2]{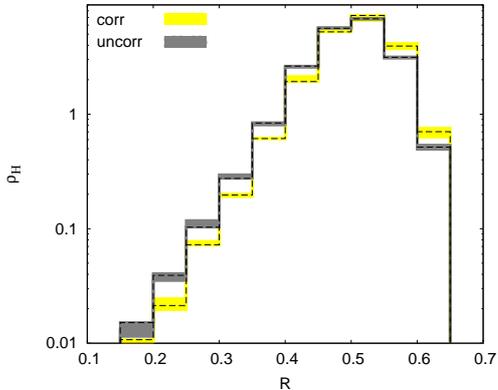}
\caption{Normalized distributions of the likelihood variable $R$ for the
spin-correlation  and no-spin-correlation  hypotheses
at the $7~{\rm TeV}$ LHC. The NLO QCD results for the distributions
are shown. The bands correspond to the choices of the
renormalization/factorization  scales $0.5\;m_t < \mu < 2 m_t$.
}
\label{fig1}
\end{center}
\end{figure}

Since  we use the leading order
\pagebreak
cross-section to compute ${\cal R}$, the following
issue appears. In general, the  NLO QCD approximation includes processes
with additional massless particles in the final state. Therefore,
we need a prescription of how to map the kinematic features of such
final states
 onto leading order kinematics.  Indeed, at leading order  the process
$ pp (p \bar p)  \to t \bar t \to b \bar b \; l \bar l \; \nu_l \bar \nu_l$
has two massless $b$-quarks.  Associating  these $b$-quarks with
two $b$-jets reconstructed according to a  well-defined jet
algorithm solves the problem of additional radiation in the event.
However, perturbatively, $b$-jets at leading order are
massless, while this
is not necessarily true in higher orders.
This feature makes it difficult to connect the leading
order kinematics that enters the calculation of ${\cal R}$ with kinematics
of the actual event.  To address this problem,  we  adopt the Ellis-Soper
jet algorithm~\cite{ellissoper}, where reconstructed  jets are always massless.

\begin{figure}[t]
\begin{center}
\includegraphics[angle=-90,scale=0.2]{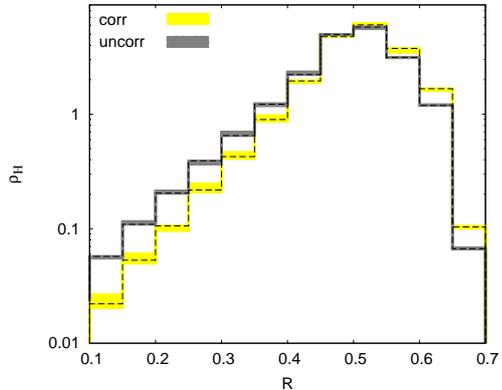}
\caption{
Normalized distributions of the likelihood variable $R$ for the
spin-correlation  and no-spin-correlation  hypotheses
at the Tevatron. The NLO QCD results for the distributions
are shown. The bands correspond to the choices of the
renormalization/factorization  scales $0.5\;m_t < \mu < 2 m_t$.
}
\label{fig2}
\end{center}
\end{figure}

The discussion in the previous paragraph tells us how
to map kinematics of a higher-order process to the kinematics
of a tree-level process.  As  input for the
calculation of the likelihood ${\cal R}({\vec x}_{\rm obs})$,
we use four-momenta of the two $b$-jets,
the four-momenta of the two charged leptons and the missing transverse
momentum, which we identify with the component of the
momentum of the two neutrinos, orthogonal to the collision
axis. We also note that, since charges of $b$-jets can not
be unambiguously defined, we require a procedure to assign one jet
to be a $b$-quark  jet and the other jet to be a $\bar b$-quark
jet. We do this
by computing the invariant mass of the positively charged lepton and the
two $b$-jets and identifying the jet that minimizes this invariant mass,
with the $b$-quark jet. The other $b$-jet is then identified with
the   $\bar b$-quark jet
and, in leading order kinematics,
 we treat this jet  as if it comes from the decay of the anti-top quark.

Having discussed a procedure to identify the input,
we turn to the calculation of ${\cal R}$; this  requires
an integration of
the tree-level differential cross-section for
$ pp (p \bar p)  \to t \bar t \to b \bar b \; l \bar l \; \nu_l \bar \nu_l$
over  unobserved components of the neutrino momentum.
In general this is difficult,  but we
assume  that the process goes through the on-shell intermediate states, so that
the invariant masses  of $b \bar l \nu$
 and $\bar b l \bar \nu$  are equal to $m_t$ and that
the invariant masses of $ \bar l \nu$ and $l \bar \nu$
are equal to  $m_W$.  Hence,
we compute
\begin{equation}
\begin{split}
&
P_{H}(p_{\rm obs},p_{\perp,\rm miss})
= {\cal N}^{-1}_H
\int
[{\rm d} p_\nu][{\rm d} p_{\bar \nu}]
\\
& \times \sum \limits_{ij}^{} f_i(x_1)f_j(x_2)|{\cal M}_H^{ij}(
p_{\rm obs},p_{\nu},p_{\bar \nu} )|^2
\\
&
\times \delta^{(2)}(\vec p_{\nu,\perp} + \vec p_{\bar \nu,\perp} - \vec p_{\perp,\rm miss})
\\
&
\times \delta(M_{\bar l \nu}^2 - m_W^2)
\delta(M_{l \bar \nu}^2 - m_W^2)
\\
&
\times
\delta(M_{\bar l \nu b}^2 - m_t^2)
\delta( M_{l \bar \nu \bar b}^2 - m_t^2),
\label{eq3}
\end{split}
\end{equation}
where  $[{\rm d} p] = {\rm d}^3 \vec p/((2\pi)^3 2 E)$ is the
invariant integration measure,
$p_{\rm obs} = \{p_{b},p_{\bar b},p_{l},p_{\bar l}\}$ is the
set of observable momenta, $f_i(x)$ are parton distribution
functions
and  $M_{ij..k}^2 = (p_i + p_j + ..+p_k)^2$ are the respective
invariant masses squared.
As we see, there are six $\delta$-functions  in Eq.(\ref{eq3}),
so that  all integration variables are fixed;
all we need to do is
to solve  the on-shell constraints. This is a standard procedure
which is described e.g. in Ref.~\cite{Sonnenschein:2006ud}; we do not repeat
such a discussion here.
In general, solving the on-shell constraints  leads to
several solutions (the maximal number is four), in which case all these
solutions should be taken into account.

The result for the
probability distribution reads
\begin{equation}
\begin{split}
& P_{H}(p_{\rm obs},p_{\perp,{\rm miss}})
 = {\cal N}^{-1}_H
\sum_{ij} \sum \limits_{a}^{}  J_a
\\
& \times
f_i^{(a)}f_j^{(a)} |{\cal M}_H^{ij,\rm LO}(
p_{\rm obs},p_{\nu}^{(a)},p_{\bar \nu}^{(a)} )|^2,
\label{eq5}
\end{split}
\end{equation}
where the second sum
is  over all the solutions that are obtained by reconstructing
the final state and $J_a$ is the Jacobian which appears when the
integration over the neutrino momentum is carried out.
Also, $f_{i,j}^{(a)} = f_i(x_{1,2}^{(a)})$ is a parton distribution
whose argument is reconstructed from the kinematics of a final state.
Finally, we emphasize that in the calculation of probability distributions
$P_{H}$
and the variable ${\cal R}$, we always use leading order matrix elements,
as explicitly shown in Eq.(\ref{eq5}).

The result for $P_H$ in Eq.(\ref{eq5}) allows us to calculate
the likelihood ${\cal R}$ and carry out the indicated program.  In practice,
however, we make use of the fact that both at the Tevatron and the LHC
there is a single partonic channel that dominates the production process.
Therefore, in
Eq.(\ref{eq5}), we use $i = (u,d),\;
j = (\bar u, \bar d)$ to compute ${\cal R}$
for the Tevatron and
$i = j = g$ to compute ${\cal R}$ for  the LHC.  We also neglect the dependence
of the normalization factor ${\cal N}_H$
on the hypothesis $H$, following the observation
that total cross-sections for $t \bar t$ pair production
are insensitive to the (non)existence of top quark
spin correlations.  Finally,
{\it in the computation of the likelihood variable}
${\cal R}({\vec x}_{\rm obs})$,
we always set the renormalization and
factorization scale to $m_t$ and use leading order parton distribution
functions.

To calculate the probability distribution of the variable
${\cal R}$
for a given hypothesis $H$,
we perform a numerical integration in Eq.(\ref{eq2}). We generate
events assuming that they must pass basic selection cuts for the
$t \bar t $ events.  For both the LHC and the Tevatron,
we require $p_{\perp}^{l} > 20~{\rm GeV}$,
$p_\perp^{\rm miss} > 40~{\rm GeV}$ and $|\eta_{l}| < 2.5$.
We also require that there are at least two $b$-jets in the event; jets
are defined using Ellis-Soper jet algorithm~\cite{ellissoper} with
$\Delta R = \sqrt{\Delta \eta^2 + \Delta \phi^2} = 0.4$ and
the  jet $p_\perp$-cut is  set to
$25~{\rm GeV}$ for both the Tevatron and the LHC.
 We use MRST2001 and MRST2004  parton distribution functions~\cite{mrst}
in
LO and NLO computations, respectively. All calculations that
we report
in this paper make use of  the numerical program for computing
NLO QCD effects in $t \bar t$ pair production,
developed in
Ref.~\cite{ms}.

We now present the results of the  calculation.
First, we compute the distribution of the likelihood variable ${\cal R}$
for both spin-correlations and for no-spin-correlations hypotheses.
These distributions   are shown in Figs.\ref{fig1},\ref{fig2} for the
LHC and the Tevatron, respectively.  The two distributions are similar
although the LHC distribution is more narrow.
Also, as follows from Figs.\ref{fig1},\ref{fig2} the scale dependence
of ${\cal R}$-distributions is small and we neglect it in what follows.

\begin{figure}[t]
\begin{center}
\vspace*{0.5cm}
\includegraphics[angle=-90,scale=0.2]{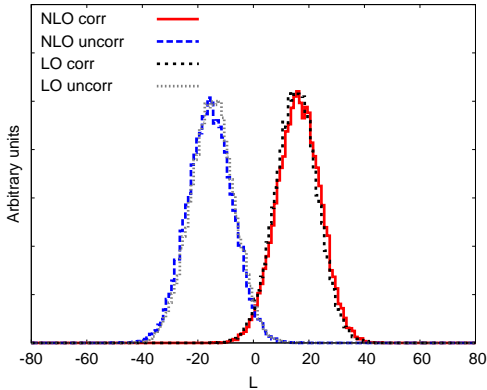}
\caption{Distributions of the likelihood ratio for correct/false
hypotheses, assuming five  hundred  dilepton events at the
$7~{\rm TeV}$ LHC.
Arbitrary units.
We show LO and NLO  QCD results  for the $L$-distribution computed
with  $\mu = m_t$.
}
\label{fig3}
\end{center}
\end{figure}

\begin{figure}[t]
\begin{center}
\vspace*{0.5cm}
\includegraphics[angle=-90,scale=0.2]{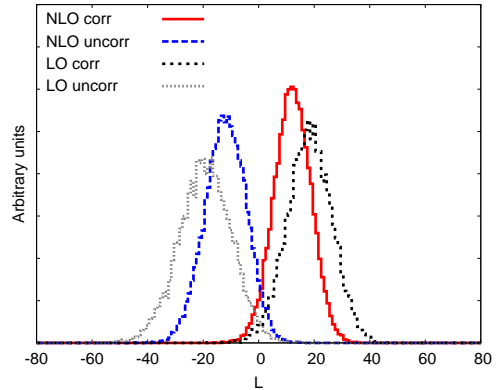}
\caption{Distributions of the likelihood ratio for correct/false
hypotheses, assuming three hundred  dilepton events at the Tevatron.
Arbitrary units.
We show  LO and NLO  QCD predictions for the $L$-distribution computed with
$\mu = m_t$.
}
\label{fig4}
\end{center}
\end{figure}

For both the Tevatron and the LHC,
there is a  difference between the two
distributions, which is especially visible
in the region of small ${\cal R}$.
To find the number of events that
is required to distinguish between the two ${\cal R}$ distributions,
we perform  a statistical test~\cite{cousins}.
To this end, we generate $N$ events
according to the probability distribution
$\rho_H({\cal R})$ defined in Eq.(\ref{eq2}) and
calculate the  quantity
\be
L = 2\; \ln [ {\cal L}_{c}/{\cal L}_{u} ],
\ee
 where
${\cal L}_K = \prod \limits_{i=1}^{N} \rho_K(R_i)$.  The statistical
interpretation of $L$ can be found in Ref.~\cite{cousins}.
We  repeat this procedure
multiple times, for $H=c,u$  and obtain two  distributions
of the variable $L$.
The distribution of the variable $L$ is peaked at positive (negative)
values if events are generated with the hypothesis $H = c$ ($H = u$)
since,  on average, ${\cal L}_c > {\cal L}_u$
(${\cal L}_u > {\cal L}_c$).  Examples of such
distributions are shown in Figs.\ref{fig3},\ref{fig4}.
To compute the   significance $S$ with which  the
hypotheses $H=c$ and $H=u$
can be separated,
we find the point beyond which the right-side tail
of the left histogram and the left-side tail of the
right histogram have equal areas.
These areas correspond to the one-sided Gaussian probability outside
of the $S/2\sigma$ range. If the two $L$-distributions  are
Gaussian with unit widths,  the significance $S$  is  the
separation between peaks of the two distributions.

The significance with which two hypotheses can be separated
depends on the number of events $N$ with which the two hypotheses
are probed.
To understand what is a  reasonable value of $N$,  we
note that  the $ pp \to t \bar t$  production  cross-section
at the $\sqrt{s} = 7~{\rm TeV}$  LHC is approximately
$160~{\rm pb}$~\cite{cms,atlas}.
Since $W$-bosons decay to electrons and muons twenty percent of the time,
and assuming thirty percent efficiency,
 we find that $1~{\rm fb}^{-1}$ of the integrated luminosity corresponds,
roughly, to   $2500$ dilepton
events. It is expected that $1~{\rm fb}^{-1}$ of luminosity will
be collected at the LHC by the end of 2011  and this sets
reasonable upper bound on the number of leptons $N$.

In fact, we do not  need that many. We take $N=500$, which corresponds
to $200~{\rm pb}^{-1}$, assuming $30\%$ efficiency. We then consider
$10^6$ pseudo-experiments and obtain the two distributions
shown in Fig.\ref{fig3}.
We convert the overlap of the two distributions into statistical significance
and
find that, with $500$ events,
 the two distributions shown in Fig.\ref{fig1} can be separated
at the $4 \sigma$ level. It is interesting to
note that the difference between NLO and LO
$L$-distributions at  the LHC is
very small, cf. Fig.\ref{fig3}.

We now turn to the discussion of the $t \bar t$ production
at the Tevatron. The production cross-section
of the $t \bar t$ pairs at the Tevatron is, approximately,  $7~{\rm pb}$
(for the latest measurements, see Refs.~\cite{d0_t,cdf_t}).
 Taking the accumulated
luminosity to be $6~{\rm fb}^{-1}$ and assuming  $30\%$ efficiency,
 we find that five hundred
dilepton ($\mu, e$) events at the Tevatron should have been observed.
We take $N = 300$ and, by considering  $10^6$ pseudoexperiments,
we obtain  $L$-distributions
shown in Fig.\ref{fig4}.
In this case, there are significant differences between $L$-distributions
computed at leading and next-to-leading order.
 Analyzing the $L$-distribution obtained with the NLO QCD approximation,
we find that,
 with $300$ Tevatron dilepton events, the spin-correlation
hypothesis can be established with the significance that
is close to $3.5\sigma$.

{\it Summary:} We have shown  that a likelihood-based analysis should
make it possible to demonstrate the existence of top quark spin correlations
in dilepton events at the Tevatron and the LHC.
We  constructed the relevant likelihood function and
computed its probability distribution through next-to-leading
order  in perturbative QCD.
Neglecting all the experimental uncertainties and the background
contributions that are relatively small for the dilepton channel, we
find that  with $500$  dilepton events at the LHC and
with $300$ dilepton events at the Tevatron the existence
of spin correlations can be established with better than
$3 \sigma$ significance. This number of events will require
just about $200~{\rm pb}^{-1}$ accumulated luminosity at the LHC and
is already available at the Tevatron.  We believe
that our  results are   sufficiently encouraging to
 warrant a more complete study including proper treatment of
experimental uncertainties and backgrounds.

{\bf Acknowledgments} We are grateful to
A. Gritsan for explaining
to us the power of statistical methods.   We acknowledge conversations
with S.~Parke that triggered our interest in physics
of top quark spin correlations. We are indebted to A. Grohsjean, S.~Parke and
Y.~Peters for discussions and for pointing out an error in the preliminary
version  of this paper.
This research is supported
by the NSF under grant PHY-0855365 and  by the start-up funds
provided by the Johns Hopkins University. Calculations reported
in this paper were performed at the Homewood High Performance Cluster
of the Johns Hopkins University.

\end{document}